\documentclass{webofc}
\usepackage[varg]{txfonts}   
\usepackage{listings}
\usepackage{hyperref}
\begin{document}
\title{Pulsating stars in $\omega$ Centauri:}
%
%
\subtitle{Near-IR properties and period-luminosity relations}

\author{\firstname{Camila} \lastname{Navarrete}\inst{1,2}\fnsep\thanks{\href{mailto:cnavarre@astro.puc.cl}{\tt cnavarre@astro.puc.cl}} \and
        \firstname{M\'arcio} \lastname{Catelan}\inst{1,2,3}\fnsep\thanks{\href{mailto:mcatelan@astro.puc.cl}{\tt mcatelan@astro.puc.cl}} \and
        \firstname{Rodrigo} \lastname{Contreras Ramos}\inst{1,2} \and
        \firstname{Javier} \lastname{Alonso-Garc\'ia} \inst{4} \and
        \firstname{Felipe} \lastname{Gran} \inst{1,2} \and
        \firstname{Istv\'an} \lastname{D\'ek\'any} \inst{5} \and
        \firstname{Dante} \lastname{Minniti} \inst{6,2,7}
}

\institute{Pontificia Universidad Cat\'{o}lica de Chile, Instituto de Astrof\'{i}sica, Av. Vicu\~{n}a Mackenna 4860,\\ 782-0436 Macul, Santiago, Chile 
\and
           Millennium Institute of Astrophysics, Santiago, Chile  
\and
	   Centro de Astro-Ingenier\'ia, Pontificia Universidad Cat\'olica de Chile, Santiago, Chile
\and
	  Unidad de Astronom\'ia, Facultad Cs. B\'asicas, Universidad de Antofagasta, Av. U. de Antofagasta 02800, Antofagasta, Chile
\and
	  Astronomisches Rechen-Institut, Zentrum f{\"u}r Astronomie der Universit{\"a}t Heidelberg, M{\"o}nchhofstr. 12-14, 69120 Heidelberg, Germany 
\and
          Departamento de Fisica, Facultad de Ciencias Exactas, Universidad Andres Bello, Av. Fernandez Concha 700, Las Condes, Santiago, Chile
\and 
             Vatican Observatory, V00120 Vatican City State, Italy
     }

\abstract{$\omega$ Centauri (NGC~5139) contains many variable stars of 
different types, including the pulsating type II Cepheids, RR Lyrae 
and SX Phoenicis stars. We carried out a deep, wide-field, near-infrared (IR)
variability survey of $\omega$~Cen, using the VISTA telescope. We assembled an 
unprecedented homogeneous and complete $J$ and $K_{\rm S}$ near-IR 
catalog of variable stars in the field of $\omega$~Cen. In this paper we compare 
optical and near-IR light curves of RR Lyrae stars, emphasizing the main 
differences. Moreover, we discuss the ability of near-IR observations to detect 
SX Phoenicis stars given the fact that the amplitudes are much smaller in these 
bands compared to the optical. Finally, we consider the case in which all the 
pulsating stars in the three different variability types follow a single  
period-luminosity relation in the near-IR bands.}
\maketitle
%


\section{Introduction}\label{sec:intro}
In the near-infrared (IR) bands, the study of variable stars has some 
advantages compared to the optical: i) the period-luminosity (PL) relations have 
lower internal dispersions (\cite{Longmore90, Catelan04}), leading to more precise 
distance determinations; ii) the interstellar extinction is lower (for the 
$K_{\rm S}$ band, the extinction is one tenth of the extinction in the $V$ 
band); and iii) pulsating stars tend to have smaller amplitudes, leading to 
better constrained mean magnitudes with a smaller number of epochs. 

Despite these advantages, only a handful of wide-field time-series surveys are 
devoted to near-IR bands. Two ongoing near-IR variability surveys are the Vista 
Variables in the V\'ia L\'actea (VVV) ESO Public Survey, which acquired time-resolved
data (in the $K_{\rm S}$ band) in highly extinct regions of the Galaxy, at low 
Galactic latitudes (see, e.g., \cite{Minniti10, Catelan2013}), and the VISTA 
survey of the Magellanic Cloud system (VMC; e.g., \cite{Cioni2011, 
Ripepi2014, Ripepi2016}). A detailed review of the time-series surveys in the 
near-IR is presented by \cite{Matsunaga17} in these proceedings.

In this present scenario, $\omega$~Centauri (NGC~5139) is an excellent 
laboratory to study pulsation properties in the near-IR bands since it hosts a 
large and rich variable star content, previously well studied using optical 
observations. In particular, the cluster has at least 180 RR Lyrae (RRL) stars 
(\cite{Navarrete15}); 74 SX Phoenicis (SX Phe), the greatest number found in a 
globular cluster (\cite{Olech05, Weldrake07}); and seven type II Cepheids (T2Cs), 
among others variability types.

\section{Observations}

Near-IR images of $\omega$~Cen were taken using the VISTA 4.1m telescope, at ESO 
Paranal Observatory \cite{Emerson10}. 42 and 100 images in $J$ and $K_{\rm S}$, 
respectively, were collected. The details of the observations and data reduction 
have been previously explained in \cite{Navarrete15}. Point-spread function 
(PSF) photometry was derived using the DAOPHOT II/ALLFRAME package 
(\cite{Stetson87, Stetson94}) for the innermost ten arcminutes region of the 
cluster, while the DoPhot package (\cite{Schechter93, Javier12}) was preferred to 
derive PSF photometry in the outer regions. 

From the observations, the intensity-average mean magnitudes for 278 previously 
known pulsating stars were recovered. Of them, the variability (i.e., sinusoidal 
light curve for the given period) was recovered for 209 pulsating stars. The 69 
stars previously known to be variables (\cite{Kaluzny04, Weldrake07}), but 
without signs of variability in the near-IR, include a subsample of low-amplitude, 
c-type RRL (RRc) stars, in addition to most of the SX~Phe stars that are known 
to be present in the cluster field.

\section{Near-IR and visible light curves of $\omega$~Cen RR Lyrae stars}\label{sec:figs-tabs}

\begin{figure*}
\centering
\includegraphics[width=13cm,clip]{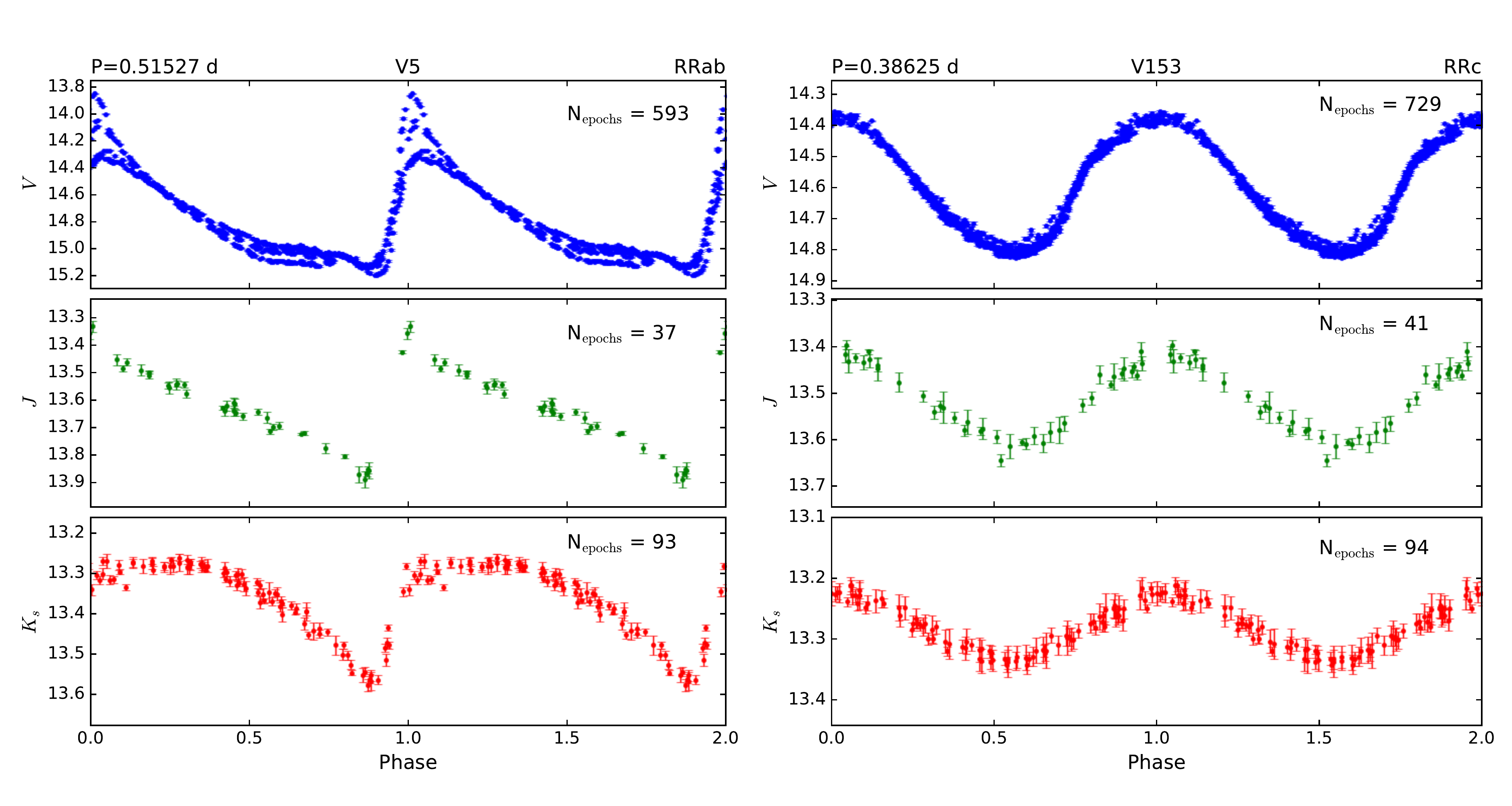}
\caption{RRL light curves in three different bands. From top to bottom, the 
light curves are in the $V$, $J$ and $K_{\rm S}$ bands, respectively. The left 
panels correspond to V5, an RRab star, while the right panels show the light 
curves for V153, an RRc star. The $V$-band data are taken from \cite{Kaluzny04}. 
For both stars, the different light curve shapes and amplitudes in different 
bandpasses are evident.}
\label{fig:grilla}
\end{figure*}

Considering the extensive optical study of variable stars in the cluster by 
\cite{Kaluzny04}, the amplitudes and light curve shapes in $V$ and the near-IR 
bands $J$ and $K_{\rm S}$ can be directly compared. Figure~\ref{fig:grilla} 
shows the light curves in $V$ (top), $J$ (middle) and $K_{\rm S}$ (bottom) for 
one ab-type RRL (RRab, left) and one RRc star (right). The light curves in $V$ 
are from the study of \cite{Kaluzny04}. Each panel shows, at the top right, the 
number of epochs in each light curve. From the figure, some striking differences 
can be noticed. First of all, the Blazhko effect that is evident in the 
$V$-band light curve of the RRab variable (V5, upper left panel) is not 
distinguishable in the near-IR light curves. This is the case for all the known 
Blazhko RRab stars in the cluster (\cite{Navarrete15}). Even though the 
resulting amplitude modulation is too small to be recovered in our 
near-IR light curves, \cite{Sollima08} 
reported increased scatter in their near-IR light curves of RR Lyr,
which they attributed to the Blazhko effect.
Second, the amplitudes in $J$ and $K_{\rm S}$ are smaller 
than in the $V$ band. \cite{Navarrete15} provide relations to convert between 
$V$, $J$ and $K_{\rm S}$ magnitudes for RRL stars, and the decrease in the 
amplitudes from the visual to the IR is evident in their Figure~3. Finally, the light curve shape 
of RRab stars tends to be more sinusoidal as the light curve is observed with 
redder filters. The same effect is seen for RRc stars but the difference is less 
prominent, given their already nearly sinusoidal $V$-band light curves. The 
different light curve shapes, as one moves from the visual regime to the near-IR, 
can be understood largely in terms of the reduced dependence of the emerging flux 
on the star's temperature variations, as compared to the radius 
(\cite{cs15}, and references therein). The lack 
of distinctive features in the light curves is one of the difficulties facing 
the automated classification of variable stars in the near-IR (see, e.g., 
\cite{Angeloni14, Elorrieta16}). 

\section{Variability amplitudes of SX Phoenicis stars}

SX~Phe stars typically have smaller variability amplitudes than RRL and T2Cs, and 
are thus more difficult to detect. $\omega$~Cen hosts 74 SX Phe, of 
which more than 45 have $V$-band amplitudes $\leq 0.1$~mag. Using the $J$ 
and $K_{\rm S}$ time series, the variability of only 12 of these SX Phe could be  
recovered. These 12 stars have the highest $V$-amplitudes among all the SX Phe in the 
cluster, from 0.13 up to 0.31 mag (based on the minimum and maximum 
magnitudes reported by \cite{Kaluzny04}). For the remaining SX~Phe stars, the 
variability amplitudes are too small and, therefore, the light curves look 
indistinguishable from those of non-variable stars. 

\begin{figure*}
\centering
\includegraphics[width=13cm,clip]{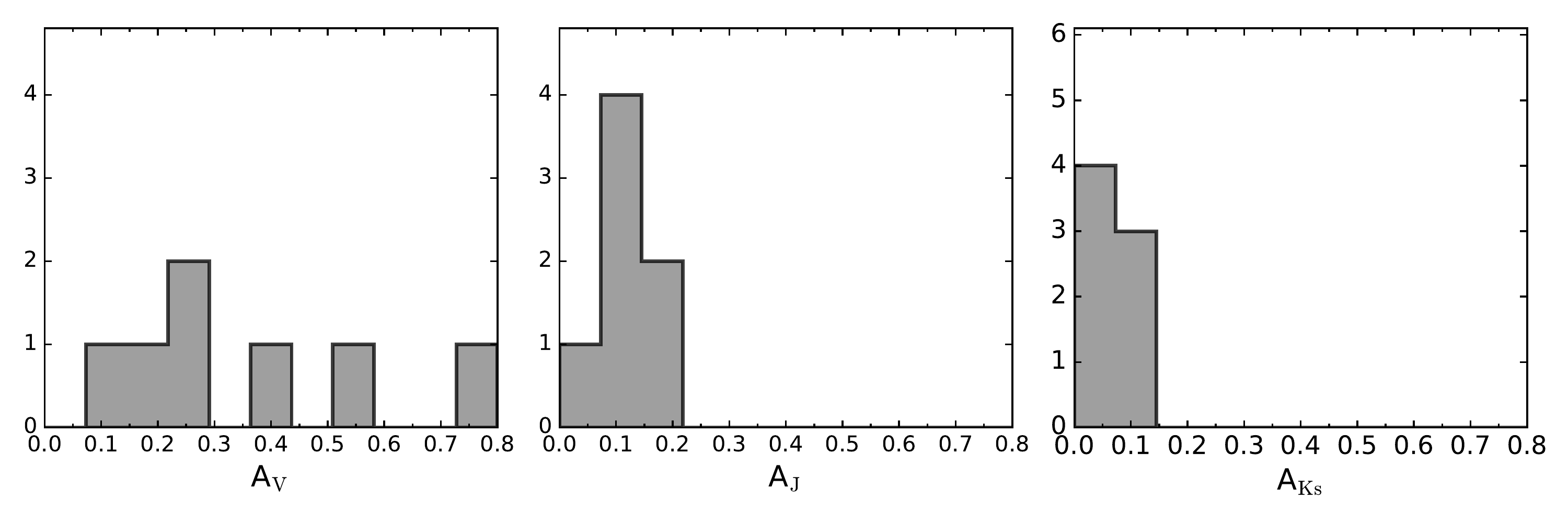}
\caption{Distribution of amplitudes for those SX Phe stars for which the 
variability was recovered in the near-IR observations and for which we were 
able to retrieve $V$-band light curves (from \cite{Kaluzny04}) and thus measure
amplitudes following the same procedure as for the near-IR light curves. 
From left to right, 
the amplitudes in $V$, $J$ and $K_{\rm S}$ bands are presented. The decrease in 
amplitude is evident when observing in IR bands.}
\label{fig:fig-3}
\end{figure*}

\begin{figure}
\centering
\sidecaption
\includegraphics[width=7cm,clip]{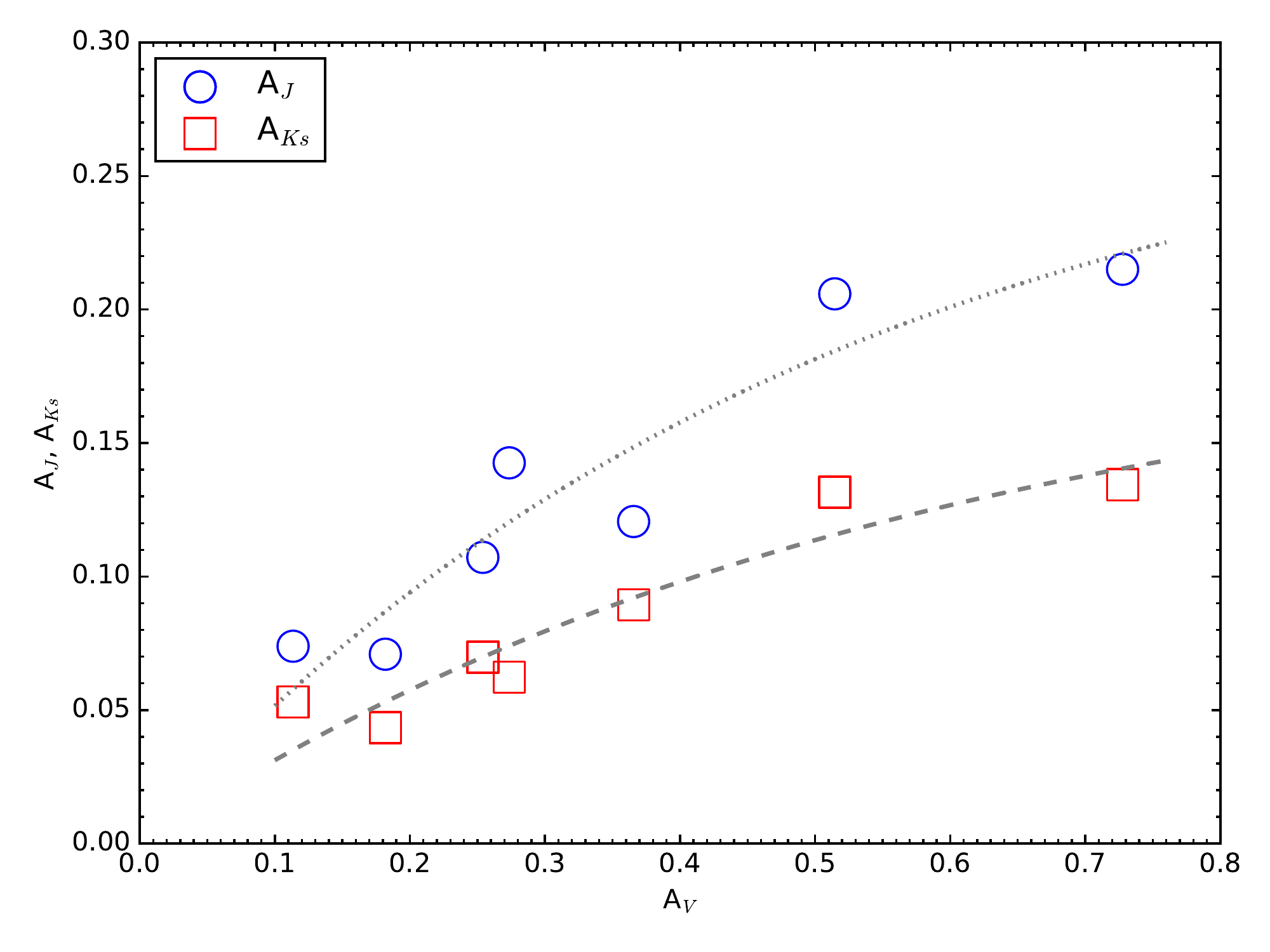}
\caption{Relationship between visual and near-IR amplitudes for SX~Phe stars
in $\omega$~Cen. The sample is the same as used in Figure~\ref{fig:fig-3}. 
$J$-band values are shown as blue circles, whereas those in the 
$K_{\rm S}$ band are displayed as red squares. The corresponding best-fitting 
relations (Eqs.~\ref{eq:ajav} and \ref{eq:akav}) are shown as 
dotted and dashed lines, respectively. 
}
\label{fig:fig-3extra}
\end{figure}

The amplitude distribution in $V$, $J$, and $K_{\rm S}$ for the SX Phe stars 
with recovered variability in the near-IR is shown in Figure~\ref{fig:fig-3}. 
In this figure, and also in Figure~\ref{fig:fig-3extra}, we include only those 
7 stars for which we were able to retrieve the corresponding $V$-band data (from 
\cite{Kaluzny04}). For these 7 stars, we re-derived the $V$-band amplitudes in 
a manner consistent with the technique that was used to obtain the near-IR 
amplitudes. 
As can be seen, most of the amplitudes in $J$ tend to be $\lesssim 0.22$~mag, 
while their respective $K_{\rm S}$ amplitudes are $\lesssim 0.15$~mag. These 
small amplitude values make the detection of SX Phe stars on the basis solely 
of near-IR observations harder. Only high-amplitude pulsators are likely to be found, 
at least at the magnitude level of $K_{\rm S} \sim 16$~mag. This group of SX Phe 
pulsators corresponds to a small fraction of the distribution of amplitudes found 
for this variability type (see, e.g., \cite{Rodriguez00}). Efforts have been 
made to collect near-IR light curves for SX Phe stars in the field as well as in 
some clusters (see, e.g., \cite{Angeloni14}), again recovering only high-amplitude 
pulsators. 

A comparison between the resulting $A_V$, $A_J$, $A_{K_{\rm S}}$ values is shown 
in Figure~\ref{fig:fig-3extra}. The lines overplotted on this diagram correspond 
to best-fit relations of the 
following forms: 

\begin{subequations}
\begin{align} 
A_J & = (0.29 \pm 0.08) \, \left[ 1 - \exp^{(-1.95 \pm 0.81) \, A_V} \right], \label{eq:ajav} \\ 
A_{K_{\rm S}} & = (0.20 \pm 0.06) \,  \left[ 1 - \exp^{(-1.74 \pm 0.76) \, A_V} \right],  \label{eq:akav}
\end{align}
\end{subequations}

\noindent which are valid for visual amplitudes in the range $A_V < 0.75$~mag. 

\section{Near-IR period-luminosity relations}

It has been suggested that the PL relations for T2Cs and RRL can be seen as a 
single relation, with nearly the same slope and zero points, at least in the near-IR, 
extending from periods shorter than 1~day up to $\sim 30$~days (see, e.g., 
\cite{Matsunaga06, Feast11, Navarrete17}). At least to first order, this may plausibly also 
extend to the SX~Phe regime (e.g., \cite{dm10,cs15}, and references therein). 
Here we revisit this possibility, using our sample of T2Cs, RRab and (candidate) 
fundamental-mode SX Phe stars from \cite{Navarrete17}. 

\begin{figure*}
\centering
\includegraphics[width=7cm,clip]{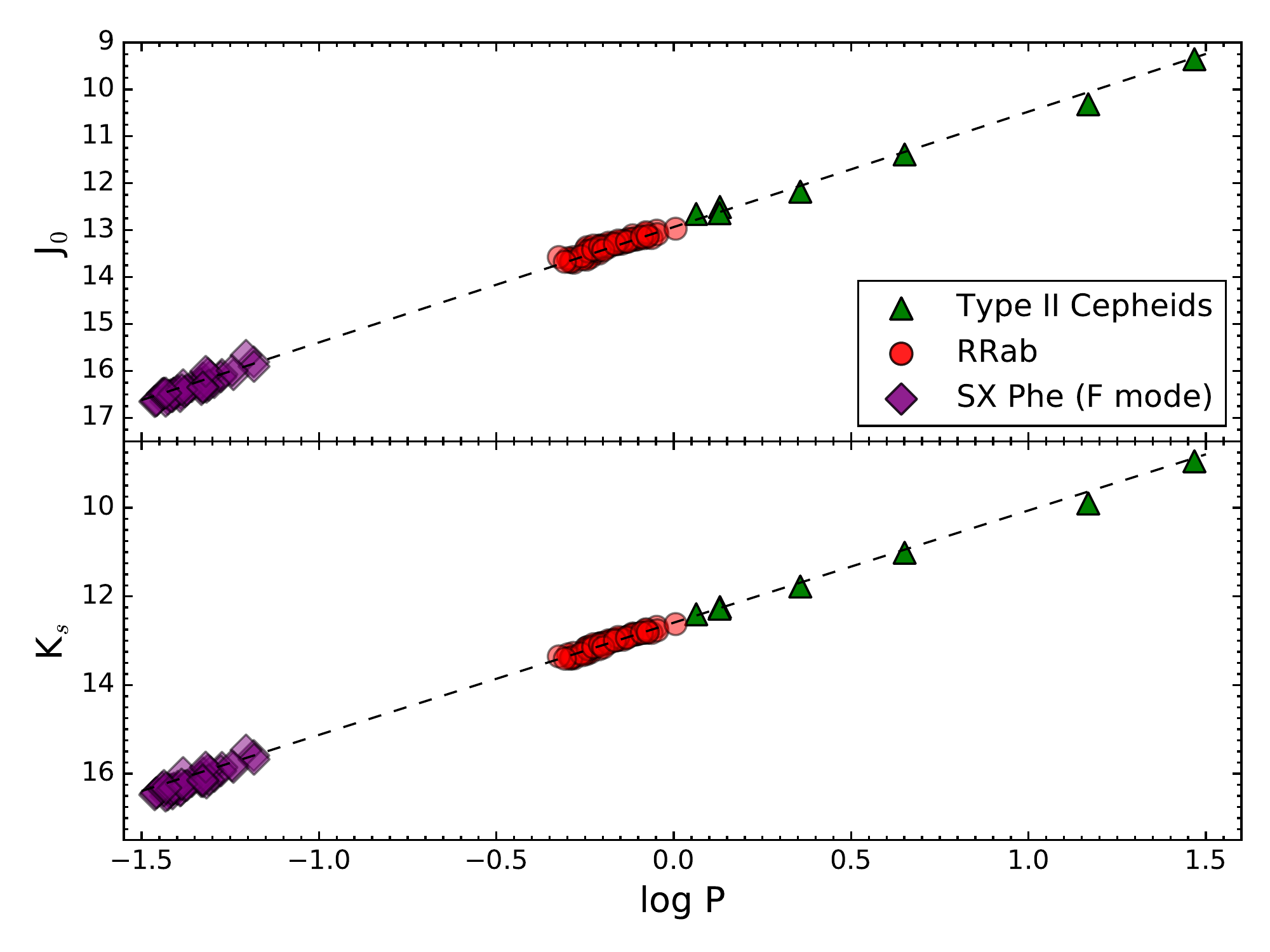}
\includegraphics[width=7cm,clip]{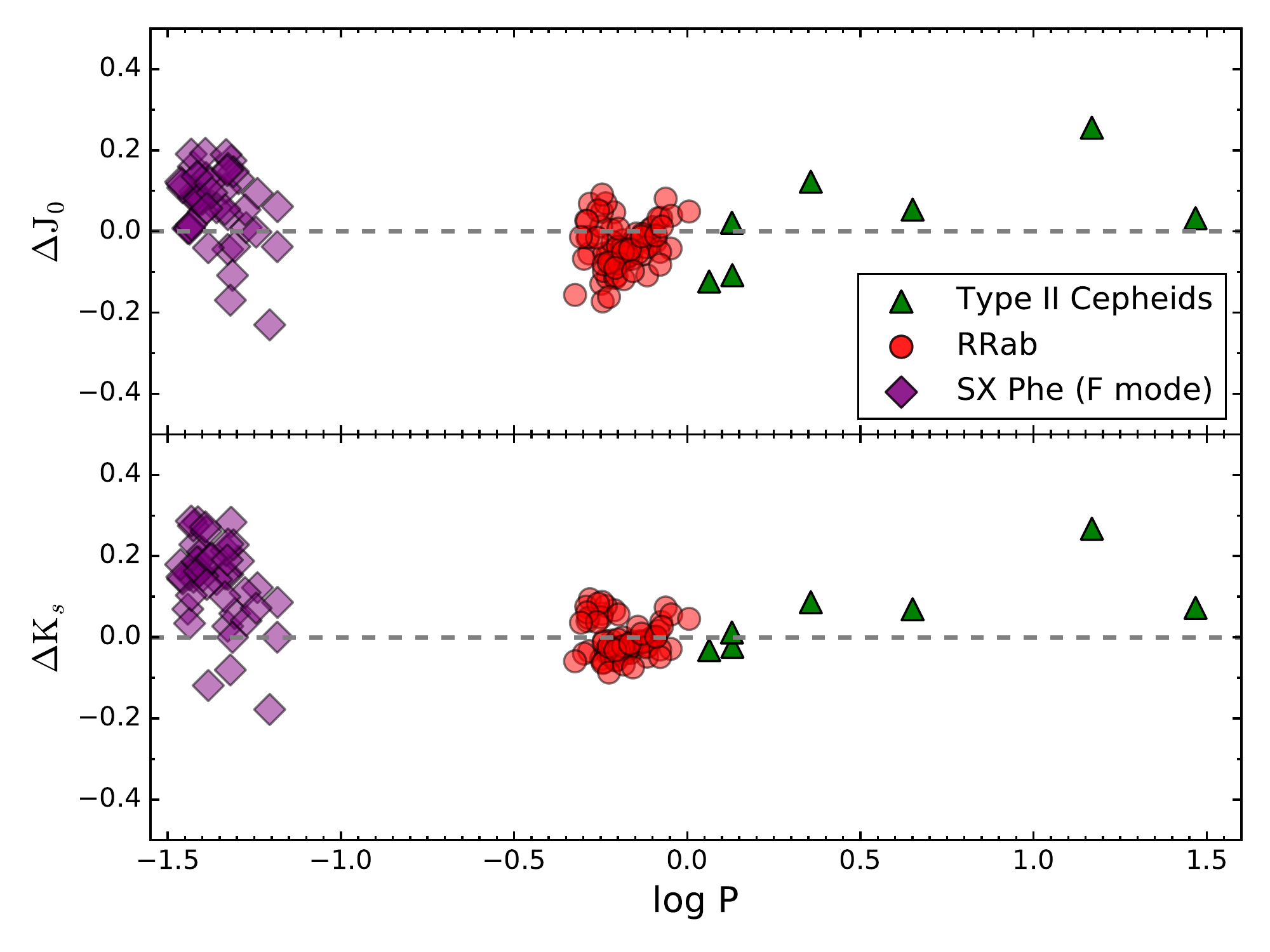}
\caption{Left: Extinction-corrected magnitudes in $J$ (top) and $K_{\rm S}$ 
(bottom) versus $\log{(P)}$, in days, for T2Cs, RRab and (candidate) fundamental-mode 
SX Phe stars. The dashed lines correspond to the linear least-squares fit to 
the data. The three variability types seem to follow, roughly, a single PL 
relation. Right: Difference between the observed extinction-corrected magnitudes 
and the derived fits, as a function of the logarithm of the period (in days). The 
horizontal dashed lines mark a zero-magnitude difference.}
\label{fig:PLall}
\end{figure*}

The left panels of 
Figure~\ref{fig:PLall} show the $\log{(P)}$ (in days) and extinction-corrected $J$ 
(top) and $K_{\rm S}$ (bottom) magnitudes and the least-squares fit recovered 
from the data (dashed lines). Despite the fact that the data largely conform to a straight line, 
resembling a single PL relation for all the aforementioned variability types, small differences 
are found, mostly in the longest- and shortest-period regimes. In the right panels of the 
figure, the difference between the observed $J$ and $K_{\rm S}$ magnitudes (top 
and bottom panels, respectively) and the best fit to the data is presented. It 
is clear that the SX Phe stars tend to deviate the most from the relation, 
suggesting a non-negligible difference in slope between their PL relation and the one 
for RRL stars. The same is found for the longest-period T2Cs (W Virginis and 
RV Tauri sub-types). RRab stars present very small deviations from the fit 
($\lesssim 0.1$~mag), in spite of the fact that its well-known metallicity 
dependence (\cite{Navarrete17}) was not corrected for when producing this plot.

\begin{table}
\centering
\caption{PL relations in different bandpasses.$^{1}$}
\begin{tabular}{lcccc}
\hline
\hline
          & $a_J$            & $b_J$ & $c_J$ & Ref. \\\hline
T2Cs      & $-2.23 \pm 0.05$ & 0 & $-0.86 \pm 0.06$ & \cite{Matsunaga06}   \\
RRL       & $-1.77 \pm 0.06$ & $0.15 \pm 0.03$ & $-0.63 \pm 0.08$ & \cite{Navarrete17}  \\
SX Phe    & $-3.04 \pm 0.17$ & 0 & $-1.60 \pm 0.22$ &  \cite{Navarrete17}  \\\hline
T2Cs+RRL$^{2}$  & $-2.32 \pm 0.03$ & 0 & $-0.77 \pm 0.06$ & This work \\
All types$^{2}$ & $-2.46 \pm 0.01$ & 0 & $-0.77 \pm 0.01$ & This work  \\
\hline
\hline
          & $a_{K_{\rm S}}$  & $b_{K_{\rm S}}$ & $c_{K_{\rm S}}$     & Ref. \\\hline
T2Cs      & $-2.41 \pm 0.05$ & 0 & $-1.11 \pm 0.05$     & \cite{Matsunaga06}   \\
RRL       & $-2.23 \pm 0.04$ & $0.141 \pm 0.02$ & $-0.96 \pm 0.05$  & \cite{Navarrete17}  \\
SX Phe    & $-3.39 \pm 0.24$ & 0 & $-2.19 \pm 0.30$     & \cite{Navarrete17}  \\\hline
T2Cs+RRL$^{2}$  & $-2.47 \pm 0.03$ & 0 & $-1.11 \pm 0.01$     & This work \\
All types$^{2}$ & $-2.53 \pm 0.02$ & 0 & $-1.12 \pm 0.01$     & This work  \\
\hline
\hline
\end{tabular}
\label{tab:tab-1}       
\\ 
\noindent $^{1}$ In the form $m_X = a_X \log{(P)} + b_X \, {\rm [Fe/H]} + c_X$, where $X$ corresponds to the bandpass.\\
\noindent $^{2}$ Adopting $(m-M)_0 = 13.708$~mag for $\omega$~Cen (from \cite{Navarrete17}).
\end{table}

The slopes of the fits can be compared with the recently derived PL and 
PL$-Z$ relations for SX~Phe and RRL stars, respectively, as obtained by 
\cite{Navarrete17} using the same data. For T2Cs, no such relations were 
derived, since the cluster hosts only seven such stars. Table~\ref{tab:tab-1} 
shows the coefficients of the PL relations for the 
different variability types (from \cite{Matsunaga06} and \cite{Navarrete17}), 
as well as 
the slopes derived in this paper simultaneously using the three 
variability types and when only T2Cs and RRL are considered. There is much better  
agreement among the different slope values when SX Phe stars are excluded from the 
fits. This was already 
evident from Figure~\ref{fig:PLall} in which SX Phe stars are those with the 
highest deviations from the fit. Reassuringly, the slope values are closer when 
the $K_{\rm S}$ band is considered, suggesting that in fact T2Cs and RRL could 
indeed share a common PL relation, at least in this bandpass.

\begin{acknowledgement} 
\noindent {\em Acknowledgments}: Support for this project is provided by the Ministry for the 
Economy, Development, and Tourism's Millennium Science Initiative through grant 
IC\,120009, awarded to the Millennium Institute of Astrophysics (MAS); by the 
Basal Center for Astrophysics and Associated Technologies (CATA) through grant 
PFB-06/2007; by CONICYT's PCI program through grant DPI20140066; and by FONDECYT 
grants \#1141141 and \#1171273 (C.N., M.C.), \#1130196 (D.M.), \#1150345 (F.G.), \#3130320 
(R.C.R.), and \#11150916 (J.A.-G.). C.N. acknowledges additional support from 
CONICYT-PCHA/Doctorado Nacional 2015-21151643.
\end{acknowledgement}


\begin{thebibliography}{}
%
%

\bibitem{Javier12}
Alonso-Garc\'ia, J., Mateo, M., Sen, B., et al. AJ, \textbf{143}, 70 (2012)
%
\bibitem{Angeloni14}
Angeloni, R., Contreras Ramos, R., Catelan, M., et al. A\&A, \textbf{567}, 100 (2014)

\bibitem{Catelan04}
Catelan, M., Pritzl, B. J., \& Smith, H. A., ApJS, \textbf{154}, 633 (2004)

\bibitem{Catelan2013}
Catelan, M., Minniti, D., Lucas, P.~W., et al., in {\em 40 Years of Variable Stars: A Celebration of Contributions by Horace A. Smith}, ed. K. Kinemuchi, C. A. Kuehn, N. De Lee, H. A. Smith, p.~139 ({\scriptsize\tt\url{https://arxiv.org/abs/1310.1996}}) (2013)

\bibitem{cs15}
Catelan, M., \& Smith, H. A., {\em Pulsating Stars} (Wiley-VCH, Weinheim, 2015)

\bibitem{Cioni2011}
Cioni, M.~-R.~L., Clementini, G., Girardi, L., et al. A\&A, \textbf{527}, 116 (2011)

\bibitem{Elorrieta16}
Elorrieta, F., Eyheramendy, S., Jord\'an, A., et al. A\&A, \textbf{595}, 82 (2016)

\bibitem{Emerson10}
Emerson, J.~P., \& Sutherland, W.~J., in \textit{Ground-based and Airborne Telescopes III.}, SPIE Conf. Ser., Vol. \textbf{7733}, ed. Stepp, L.~M., Gilmozzi, R., Hall, H.~J. (2010)

\bibitem{Feast11}
Feast, M., in {\em RR Lyrae Stars, Metal-Poor Stars, and the Galaxy}, ed. A. McWilliam, Carnegie Observatory Astrophysics Series, \textbf{5}, 170 (2011)

\bibitem{Longmore90}
Longmore, A. J., Dixon, R., Skillen, I., et al., MNRAS, \textbf{247}, 684 (1990)

\bibitem{Kaluzny04}
Kaluzny, J., Olech, A., Thompson, I. B., et al., A\&A, \textbf{424}, 1101 (2004)

\bibitem{dm10}
Majaess, D. J., JAAVSO, \textbf{38}, 100

\bibitem{Matsunaga06}
Matsunaga, N., Fukushi, H., Nakada, Y., et al., MNRAS, \textbf{370}, 1979 (2006)

\bibitem{Matsunaga17}
Matsunaga, N., preprint ({\scriptsize\tt\url{https://arxiv.org/abs/1705.02547}}) (2017)

\bibitem{Minniti10}
Minniti, D., Lucas, P.~W., Emerson, J.~P., et al., New Astron., \textbf{15}, 433 (2010)

\bibitem{Navarrete15}
Navarrete, C., Contreras Ramos, R., Catelan, M., et al., A\&A, \textbf{577}, 99 (2015)

\bibitem{Navarrete17}
Navarrete, C., Catelan, M., Contreras Ramos, R., et al., A\&A, in press ({\scriptsize\tt\url{https://arxiv.org/abs/1704.0303}}) (2017)

\bibitem{Olech05}
Olech, A., Dziembowski, W. A., Pamyatnykh, A. A., et al., MNRAS, \textbf{363}, 40 (2005)

\bibitem{Ripepi2014}
Ripepi, V., Marconi, M., Moretti, M.~I., et al., MNRAS, \textbf{437}, 2307 (2014)

\bibitem{Ripepi2016}
Ripepi, V., Marconi, M., Moretti, M. I., et al., ApJS, \textbf{224}, 21 (2016)

\bibitem{Rodriguez00}
Rodr\'iguez, E., \& L\'opez-Gonz\'alez, M.~J., A\&A, \textbf{359}, 597 (2000)

\bibitem{Schechter93}
Schechter, P.~L., Mateo, M., \& Saha, A., PASP, \textbf{105}, 1342 (1993)

\bibitem{Sollima08}
Sollima, A., Cacciari, C., Arkharov, A. A. H., et al., MNRAS, \textbf{384}, 1583 (2008)

\bibitem{Stetson87}
Stetson, P.~B., PASP, \textbf{99}, 197 (1987)

\bibitem{Stetson94}
Stetson, P.~B., PASP, \textbf{106}, 250 (1994)

\bibitem{Weldrake07}
Weldrake, D. T. F., Sackett, P. D., \& Bridges, T. J., AJ, \textbf{133}, 1447 (2007)

\end{thebibliography}
%
%

\end{document}